\documentstyle[12pt]{article}
\newcommand{\be}{\begin{eqnarray}}
\newcommand{\ee}{\end{eqnarray}}
\newcommand{\half}{{\textstyle\frac{1}{2}}}
\newcommand{\quarter}{{\textstyle\frac{1}{4}}}

\newcommand{\Pslash}{P\hspace{-.5em}/\hspace{.15em}}
\newcommand{\PPi}{P_{\rm i}}
\newcommand{\PPf}{P_{\rm f}}
\newcommand{\fourint}[1]{\int\!\frac{d^4 #1}{(2\pi)^4}}
\newcommand{\cutint}[1]{\int_{\Lambda_B}\!\frac{d^4 #1}{(2\pi)^4}}
\begin{document}
%
%
\rightline{UNITUE--THEP--2/95}
\rightline{hep-ph/9502217}
\rightline{January 1995}
\vspace{.5cm}
\begin{center}
\begin{large}
{\bf Baryon form factors in a diquark--quark bound state
description$^\dagger$} \\
\end{large}
\vspace{1cm}
{\bf G. Hellstern}$^{\rm 1}$ \\
\vspace{0.2cm}
{\em Institut f\"ur Theoretische Physik \\
Universit\"at T\"ubingen \\
Auf der Morgenstelle 14 \\
D--72076 T\"ubingen, Germany} \\[.5cm]
{\bf C. Weiss}$^{\rm 2}$ \\
\vspace{0.2cm}
{\em Institut f\"ur Theoretische Physik II \\
Ruhr--Universit\"at Bochum \\
D--44780 Bochum, Germany}
\end{center}
\vspace{1cm}
\begin{abstract}
\noindent
Nucleon form factors are calculated in a
relativistic diquark--quark picture based on the Nambu--Jona-Lasinio model.
The nucleon wave function is obtained in a static approximation
to the quark exchange interaction between the valence quark and the
diquark. We evaluate the valence quark and $0^+$--diquark contribution to
the nucleon electromagnetic and weak currents. We find reasonable
electric charge radii, magnetic moments as in the additive diquark model,
and $g_A \approx 1$. We discuss the dependence on the model parameters.
\end{abstract}
\vfill
\rule{5cm}{.15mm}
\\
\noindent
{\footnotesize $^\dagger$ Supported by COSY under contract 41170833} \\
{\footnotesize $^{\rm 1}$ E-mail:
hellst@ptdec1.tphys.physik.uni-tuebingen.de} \\
{\footnotesize $^{\rm 2}$ E-mail: weiss@hadron.tp2.ruhr-uni-bochum.de} \\
\newpage
%
%
%
A quantitative understanding of baryon form factors at low energies in
terms of quark degrees of freedom remains a challenging problem. Attempts
to base a low--energy description of baryons on QCD usually rely on the
limit $N_{\rm C}\rightarrow\infty$, in which a soliton picture emerges.
While properly describing the effects of the non--perturbative meson
cloud, this approach suffers from difficulties related to the breaking
of translational invariance. For many purposes, such as {\em e.g.}\ the
calculation of timelike nucleon form factors, it is important to
maintain proper relativistic kinematics. A description of the nucleon
as a relativistic bound state can be realized in quark models with an
effective 2--body interaction \cite{cahill_89,reinhardt_90,kaschluhn_92}.
For $N_{\rm C} = 3$, this naturally leads to the concept of diquarks as
effective constituents \cite{anselmino_93}.
The simplest such model is a Nambu--Jona-Lasinio (NJL--) model with
a local (current--current) interaction in the $\bar 3$--color channel.
A corresponding color--singlet interaction describes the spontaneous
breaking of chiral symmetry and provides a reasonable account of
meson properties \cite{ebert_86,lutz_90}. The NJL model can be rewritten
as an effective theory of mesons and baryons \cite{reinhardt_90}, in which
bound states of composite diquarks and quarks occur due to quark exchange.
The spectrum of $\frac{1}{2}^+$--baryons has been calculated using a static
approximation for the quark exchange interaction \cite{buck_92}. The Faddeev
equation for the nucleon has also been solved
exactly \cite{bentz1_93}, including also the $1^+$--diquark channel
\cite{bentz2_93,meyer_94}.
\par
Here, we apply this description to the study of the nucleon
form factors at low momentum. We consider the simplest case of a bound
state of a $0^+$--diquark and a valence quark and take the baryon wave
functions in the static approximation to the quark exchange.
Given the complexity of any relativistic description of baryons, it is
useful to first explore this simple approximation.
The present study extends a previous analysis of the form factors
in an additive diquark--quark scheme \cite{weiss_93}. Contrary to that
approach, we now derive the relativistic nucleon wave function within
the model quark dynamics in a crude but consistent approximation. Our
intention is to estimate the role of diquark and valence quark currents in
the nucleon electromagnetic and weak form factors. Moreover, we wish to
point out some general features of the bound state description of baryons,
which are likely to persist in more elaborate calculations with exact
Faddeev wave functions.
\par
The basis of our approach is the effective baryon action, which is
obtained by rewriting the partition function of the NJL--model
in terms of composite diquark and baryon
fields \cite{reinhardt_90,kaschluhn_92}. It is defined in terms of
the baryon Greens function,
\be
G_B &=& [G_0^{-1} - H]^{-1} + \Delta G_B .
\label{G_B}
\ee
Here,
\be
G_0 (k, k'; P) &=& {\cal D}(\half P - k) G(\half P + k' )
\delta (k - k' )
\ee
is the product of the scalar diquark propagator, ${\cal D}$, which is the
solution of the diquark Bethe--Salpeter equation of the NJL
model \cite{reinhardt_90}, and the quark propagator,
$G(k) = (\rlap / k - M)^{-1}$. We consider the isospin limit,
$M_u = M_d \equiv M$. $P$ and $k$ are the total and relative four--momentum
of the system. Furthermore,
\be
H(k, k' ) &=& \Gamma G(-k - k' ) \Gamma
\ee
is the quark exchange interaction. For a scalar diquark,
$\Gamma = i\gamma_5 C \epsilon$, where $C$ is the charge conjugation matrix
and $\epsilon$ the generator of the $\bar 3$--representation of the color
group. The contact term in eq.(\ref{G_B}), $\Delta G_B$, will not be
important in the following. Baryon masses and bound state wave functions are
obtained as color--singlet solutions of the Faddeev--type equation
\be
\fourint{k'} G^{-1}_B (k, k' ; P) \psi (k' | P) &=& 0, \hspace{2em}
P^2 = m_B^2 .
\label{faddeev_eq}
\ee
In the static approximation for the quark exchange \cite{buck_92},
\be
H(k, k' ) &\rightarrow& -\Gamma \frac{1}{M} \Gamma ,
\label{static_approx}
\ee
the baryon wave function in eq.(\ref{faddeev_eq}) can be written as
\be
\psi (k | P) &=& {\cal D} (\half P - k) G(\half P + k) \phi (P) .
\ee
The baryon amplitude, $\phi (P)$, is then obtained as the solution
of a Dirac--type equation,
\be
\left( A(P^2 ) \Pslash + B(P^2 ) - M \right) \phi (P) &=& 0,
\label{dirac_equation}
\ee
where the effective Dirac operator is defined as
\be
A(P^2 ) \Pslash + B(P^2 ) &=& \cutint{k} {\cal D}(\half P - k)
G(\half P + k) .
\label{dirac_operator}
\ee
In this approximation the baryon amplitude does not depend on the relative
momentum of the diquark--quark pair. The integral
in eq.(\ref{dirac_operator}) is defined with a momentum cutoff,
$\Lambda_B$, which effectively controls the width of the baryon
wave function. ($\Lambda_B$ is not identical to the NJL cutoff, {\em cf.}\
below.) We normalize the on--shell baryon amplitude by the condition
\be
\bar\phi (P) \left[ P^\mu \frac{\partial}{\partial P_\mu} \left(
A(P^2 ) \Pslash + B(P^2 )\right) \right]_{P^2 = m_B^2} \phi (P) &=& m_B .
\label{normalization}
\ee
Eq.(\ref{normalization}) fixes the correct relativistic dispersion
law for the c.o.m.\ motion of the bound state and is identical to the
usual normalization condition for Bethe--Salpeter amplitudes.
\par
Baryon form factors are calculated by coupling to the free baryon action
an electromagnetic or weak gauge field, ${\cal A}_\mu$, by minimal
substitution at quark level. One then obtains the current of the baryon field
as the corresponding functional derivative of the inverse baryon
propagator, eq.(\ref{G_B}). The matrix element of the baryon electromagnetic
current for a transition $\PPi \rightarrow \PPf$, where
$\PPi^2 = \PPf^2 = m_B^2$, is in general given by
\be
J_B^\mu (\PPf , \PPi ) &=& \fourint{k}\fourint{k'}
\,\bar\psi (k|\PPf )\, \frac{\delta G_B^{-1}}{\delta {\cal A}^\mu}
(k,k'; \PPf , \PPi ) \, \psi (k'|\PPi )
\label{Gamma_B} \\
&=& \left( F_B^{\rm e} (Q^2 ) - F_B^{\rm m} (Q^2 ) \right)
2 m_B \frac{(\PPf + \PPi )^\mu}{(\PPf + \PPi )^2}
+ F_B^{\rm m} (Q^2 ) \gamma^\mu .
\ee
Here, $Q = \PPf - \PPi$, and $F_B^{\rm e} , F_B^{\rm m}$ are the baryon
electric and magnetic form factor. The baryon current consists of terms
describing the coupling of the external field to the diquark propagator, the
quark propagator and the quark exchange. (The contact term in eq.(\ref{G_B})
does not contribute to the baryon current on the mass shell.) In the static
approximation, eq.(\ref{static_approx}), it is consistent to neglect the
quark exchange current. Note that the quark exchange also does not
contribute to the normalization of the baryon amplitude, {\em cf.}\
eq.(\ref{normalization}). Thus, in the static approximation,
eq.(\ref{Gamma_B}) becomes
\be
J^\mu_B (\PPf , \PPi ) \!\! &=& \!\! \bar\phi (\PPf ) \left(
\Gamma^\mu_{B, {\rm diq}} (\PPf , \PPi )
+ \Gamma^\mu_{B, {\rm quark}} (\PPf , \PPi )
+ C(\Lambda_B ) \gamma^\mu \right) \phi (\PPi )
\label{Gamma_B_static} \\
\Gamma^\mu_{B, {\rm diq}} (\PPf , \PPi ) \!\! &=& \!\! \cutint{k}
{\cal D}(\half\PPf - k) {\cal F}^\mu (\half\PPf - k, \half\PPi - k)
{\cal D}(\half\PPi - k) G(\quarter\PPf + \quarter\PPi + k) \nonumber \\
&& \label{J_diq} \\
\Gamma^\mu_{B, {\rm quark}} (\PPf , \PPi ) \!\! &=& \!\! \cutint{k}
{\cal D}(\quarter\PPf + \quarter\PPi - k)
G(\half\PPf + k) \Gamma^\mu_q G(\half\PPi + k)
\label{J_quark}
\ee
We note that the on--shell baryon current defined by
eqs.(\ref{J_diq}, \ref{J_quark}) is transverse. Here,
\be
{\cal F}^\mu (k, k') &=& (Q_u + Q_d ) Z_{0^+}^{-1} \left\{
(k + k')^\mu F_{0^+} (q^2, k^2 , k^{'2})
+ q^\mu G_{0^+}(q^2, k^2 , k^{'2}) \right\} ,
\label{F_diquark}
\ee
with $q = k - k'$, is the off--shell e.m.\ vertex function of the extended
$0^+$--$ud$--diquark \cite{weiss_93,weiss_94}. The valence quark is taken as
pointlike, $\Gamma^\mu_q = i Q_{u, d} \gamma^\mu$, corresponding,
respectively, to a proton or a neutron.
($Q_{u, d} = \frac{2}{3}, -\frac{1}{3}$ are the quark charges.) In the
approach based on \cite{reinhardt_90}, a
valence quark form factor would arise if dynamical meson exchange were
taken into account\footnote{A consistent treatment of meson exchange effects
would require the inclusion of meson exchange in the quark--quark interaction
and the diquark--quark binding}. Furthermore, we have included in
eq.(\ref{Gamma_B_static}) a counter term, $C(\Lambda_B )\gamma^\mu$,
which we choose to fix charge conservation at $Q^2 = 0$. This is necessary
when working with a momentum--space cutoff. However, in the numerical
calculations we find that the violation of charge
conservation is small, {\em i.e.}, $C(\Lambda_B ) \ll 1$. Note also that
the counter term does not affect the charge radii.
\par
The composite diquark propagator and the corresponding e.m.\ vertex function,
eq.(\ref{F_diquark}) have been determined within the gauge invariant
proper--time regularized NJL model \cite{weiss_93,weiss_94}. To facilitate
the evaluation of the loop integrals in eqs.(\ref{J_diq}, \ref{J_quark}), we
approximate the full diquark propagator by its pole form,
\be
{\cal D}(k) &\rightarrow& \frac{Z_{0^+}}{k^2 - m_{0^+}^2} ,
\label{D_pole}
\ee
where the pole and the residue are determined from the exact solution.
This is justified if the cutoff defining the baryon wave function,
$\Lambda_B$, is not too large, which is the case in the calculations below.
Consistently with the pole form of the propagator, eq.(\ref{D_pole}), we make
an on--shell approximation\footnote{Note that the off--shell dependence of
the exact ${\cal D}(k)$ and ${\cal F}^\mu (k, k')$ cancels to leading order
in $(k^2 - m_d^2)$ and $(k^{'2} - m_d^2)$, when considering the product
${\cal D}(k) {\cal F}^\mu (k, k' ) {\cal D}(k' )$ \cite{weiss_94}.} for the
composite diquark vertex function, {\em i.e.}, we replace in
eq.(\ref{F_diquark}) the diquark form factor $F_{0^+}(q^2 , k^2 , k^{'2})$
by its value at $k^2 = k^{'2} = m_{0^+}^2$. (The longitudinal part of
eq.(\ref{F_diquark}) does not contribute to the on--shell baryon form
factor.) For regularization of the diquark--quark loop integrals in
eqs.(\ref{J_diq}, \ref{J_quark}) we employ a sharp euclidean cutoff,
$\Lambda_B$, after introducing Feynman parameters, {\em cf.}\
\cite{lutz_90}. (We have also done calculations with a proper--time--type
regularization for the diquark--quark loop and found only very small
differences to cutoff regularization.) The contributions of
eqs.(\ref{J_diq}, \ref{J_quark}) to the nucleon electric and magnetic
form factor can then be identified after making use of the Dirac equation,
eq.(\ref{dirac_equation}), to simplify the numerators of the Feynman
integrals.
\par
Before discussing results for the baryon charge radii and magnetic moments,
it is instructive to consider the dependence of these quantities on the
baryon c.o.m.\ momentum. Fig.1 shows, for a fixed quark and diquark mass,
the proton charge radius from eq.(\ref{Gamma_B_static}) as a function of
$P^2 \equiv \PPi^2 = \PPf^2 \neq m_B^2$. Here, the amplitudes are
normalized according to eq.(\ref{normalization}), but at $P^2\neq m_B^2$.
(One may imagine the off--shell values of $P^2$ to be generated by varying
the strength of the static quark exchange term, $M$, in
eq.(\ref{dirac_equation}).)
The charge radius grows and eventually diverges, as $P^2$ approaches the
diquark--quark continuum threshold. The presence of a singularity is, of
course, due to the lack of confinement of this model. The strong variation
of $\langle r^2 \rangle_p$ with $P^2$ for bound baryons, however, should not
be regarded as unphysical. We know that even in confining theories there
are ``would--be''--thresholds related to losely bound states \cite{jaffe_92}.
In this spirit, fig.1 simply reflects the fact that, in the context of the
effective quark dynamics described by the NJL model, the baryon has to be
considered as a losely bound state, contrary to {\em e.g.}\ the pion. This
is in agreement with the fact that the diquark masses obtained in the NJL
model \cite{vogl_91} as well as in the instanton vacuum \cite{schaefer_94}
are typically of the order of $0.5\ldots 0.6\,{\rm GeV}$, so that the
diquark--quark threshold is close to the physical nucleon mass. Furthermore,
the exact Faddeev calculation leads to nucleon binding energies of
$\varepsilon_B = M + m_{0^+} - m_B \approx 50\ldots 100\,{\rm MeV} \ll m_B$
\cite{bentz1_93}. When fixing the parameter $\Lambda_B$ of the static
approximation, we  must keep in mind that it is the binding energy --- not
the absolute mass --- which is smoothly related to the characteristics of
the baryon wave function. We thus take values of $\Lambda_B$,
for which the Dirac equation, eq.(\ref{dirac_equation}) gives binding
energies comparable to those of the exact solutions (see fig.1). These
lead to realistic charge radii. We see no physical
grounds for identifying $\Lambda_B$ with the NJL cutoff employed in the
diquark and meson sector \cite{buck_92}, the more since values of
$\Lambda_B \approx 0.6\ldots 0.7\,{\rm GeV}$
would considerably overestimate the binding energy, {\em cf.}\ fig.1
and \cite{bentz1_93}. For a given value of $\Lambda_B$, we then choose
the NJL coupling constant in the diquark channel such as to reproduce the
physical nucleon mass in eq.(\ref{dirac_equation}) \cite{buck_92}.
We emphasize that, in this way, the baryon form factors
are not sensitive to the precise value of the baryon mass, if
$\Lambda_B$ (and thus, approximately, $\varepsilon_B$) is kept fixed.
\par
With the parameters of the model determined in this way, we may now discuss
the results for the on--shell nucleon charge radii and magnetic moments.
As expected, the proton charge radius depends inversely
on the cutoff used in the diquark--quark wave function. The experimental
radius, $\langle r^2\rangle_p = 0.74\,{\rm fm}^2$, can be reproduced for
$\Lambda_B \approx 0.32\,{\rm GeV}$. Since this value for $\Lambda_B$ would
probably be larger, if the $1^+$--diquarks were included \cite{weiss_93}, we
give in table 1 also results for larger values of $\Lambda_B$. In addition,
we have performed calculations with a pointlike diquark form factor,
$F_{0^+} \equiv 1$,
{\em cf.}\ eq.(\ref{F_diquark}). Comparison of the values obtained with
extended and pointlike scalar diquarks shows that the intrinsic charge
radius of the scalar diquark contributes about $10\ldots 20\%$ of the
proton charge radius. The neutron charge radius is reproduced well.
This quantity originally provided evidence for scalar diquark
correlations \cite{dziembowski_81}. The intrinsic diquark
radius somewhat increases the neutron charge radius, improving the agreement
with the experimental value. The magnetic moments are
not too different from their values in an additive diquark--quark
picture with only $0^+$--diquarks, $\mu_p = \frac{2}{3}(e/2M),
\mu_n = -\frac{1}{3}(e/2M)$. Here, large contributions are
to be expected from the $1^+$--diquark channel. Contrary to the charge
radii, the magnetic moments are rather independent of the size of the
baryon wave function.
\par
We have also calculated the isovector axial coupling constant of the
nucleon\footnote{We do not address here questions related to the
induced pseudoscalar form factor and the Goldberger--Treiman
relation, which require the inclusion of meson fluctuations, but
stay within a valence quark picture. See \cite{ericson_88} for a discussion
of these issues.}, $g_A$. In this
case the quark vertex is given by $\Gamma^{5\mu}_q = i \tau^a
\gamma_5\gamma^\mu$.
The scalar diquark has no axial coupling, so that the baryon axial form
factor is determined entirely by the corresponding valence quark axial
current, eq.(\ref{J_quark}). (We do not include a counter term in
eq.(\ref{Gamma_B_static}) now.) As shown in table 1, for realistic
parameters, $g_A < 1$. The value $g_A = 1$ would be obtained in
an additive diquark--quark scheme with only $0^+$--diquarks.
Corrections to this value are seen to increase with $\Lambda_B$, as the
diquark--quark system becomes more strongly bound. Note that
also in $g_A$ we expect sizable contributions from the $1^+$--diquark
channel, both through the axial coupling of the $1^+$--diquark as well
as through $0^+$--$1^+$ transitions.
\par
In summary, our results show two types of behavior of nucleon
observables. Charge radii depend strongly on the extent of the
diquark--quark wave function and can reasonably be described in a picture
with (even pointlike) $0^+$--diquarks. On the contrary, magnetic moments
and $g_A$ are rather insensitive to the strength of the diquark--quark
binding and therefore amenable to an additive diquark--quark
description \cite{weiss_93}. They require, however, the inclusion of the
$1^+$--diquarks.
\par
In our simple estimates, we have treated the cutoff, $\Lambda_B$, as a
parameter and related it to the proton charge radius. In principle,
the width of the diquark--quark wave function
should be determined by the underlying effective quark dynamics.
It remains to be seen whether the quark exchange mechanism included
in eq.(\ref{faddeev_eq}) leads by itself to realistic baryon radii,
or whether these quantities are dominated by other effects
(dynamical meson exchange, confinement). Meanwhile, it is of practical
interest that, with appropriate values of $\Lambda_B$, the static
approximation seems to offer the possibility of a reasonable approximation
to the nucleon wave function, if $1^+$--diquarks are included. It may
{\em e.g.}\ be employed to calculate strong form factors of the nucleon.
\par
The authors are grateful to H.\ Reinhardt and A.\ Buck for helpful
discussions.
\newpage
%
%

%
\newpage
%
%
\noindent
\begin{table}[t]
\centering
\[
\begin{array}{|r|c|c|c|c|}
\hline
& \Lambda_B = 0.323\,{\rm GeV} & \Lambda_B = 0.35\,{\rm GeV} &
\Lambda_B = 0.45\,{\rm GeV} & {\rm exp.} \\
\hline
g_2 / g_1 & 1.46 & 1.42 &  1.27 & \\
m_{0^+}/{\rm GeV} & 0.582 & 0.593 & 0.631 & \\
Z_{0^+}^{1/2} & 14.3 & 14.0 & 13.0 & \\
\varepsilon_B /{\rm GeV} & 0.044 & 0.055 & 0.093 & \\
\langle r^2\rangle_p /{\rm fm}^2 & 0.74\; (0.66) & 0.65\; (0.57) &
0.48\; (0.40) & 0.74 \\
\langle r^2\rangle_n /{\rm fm}^2 & -0.11\; (-0.19) & -0.10\; (-0.17) &
-0.07\; (-0.15) & -0.12 \\
\mu_p / (e/2m_B ) &  1.67 &  1.67 &  1.75  & 2.79 \\
\mu_n / (e/2m_B ) & -0.78 & -0.82 & -0.83  & -1.91 \\
g_A               & 0.96  &  0.95 &  0.91  & 1.26 \\
\hline
\end{array}
\]
\caption[]{\it The nucleon electric charge radii,
$\langle r^2 \rangle_{p, n}$, magnetic moments, $\mu_{p, n}$, and isovector
axial coupling constant, $g_A$, for various values of the cutoff in the
diquark--quark momentum, $\Lambda_B$. Results are given for a constituent
quark mass of $M = 0.4\,{\rm GeV}$. For each value of $\Lambda_B$, the
coupling constant of the scalar diquark channel, $g_2$, has been chosen to
fit the physical nucleon mass, $m_B = m_p = 0.938\,{\rm GeV}$. The
corresponding values of $m_{0^+}, Z_{0^+}$, and the diquark form factor have
been determined from the full diquark propagator of the proper--time
regularized NJL model {\rm \cite{weiss_93}}. The ratio of the diquark
coupling constant to that in the meson channel, $g_2 / g_1$, is as defined
in {\rm \cite{buck_92,weiss_93}}. Also given is the baryon binding
energy, $\varepsilon_B = M + m_{0^+} - m_p$. With a value of
$\Lambda_B = 0.323\,{\rm GeV}$, the experimental proton charge radius can
be reproduced. The charge radii in parentheses have been calculated with a
pointlike diquark form factor, $F_{0^+} \equiv 1$,
cf.\ {\rm eq.(\ref{F_diquark}).}}
\end{table}
\newpage
%
%
\noindent {\bf\large Figure caption}
\vspace{1cm} \\
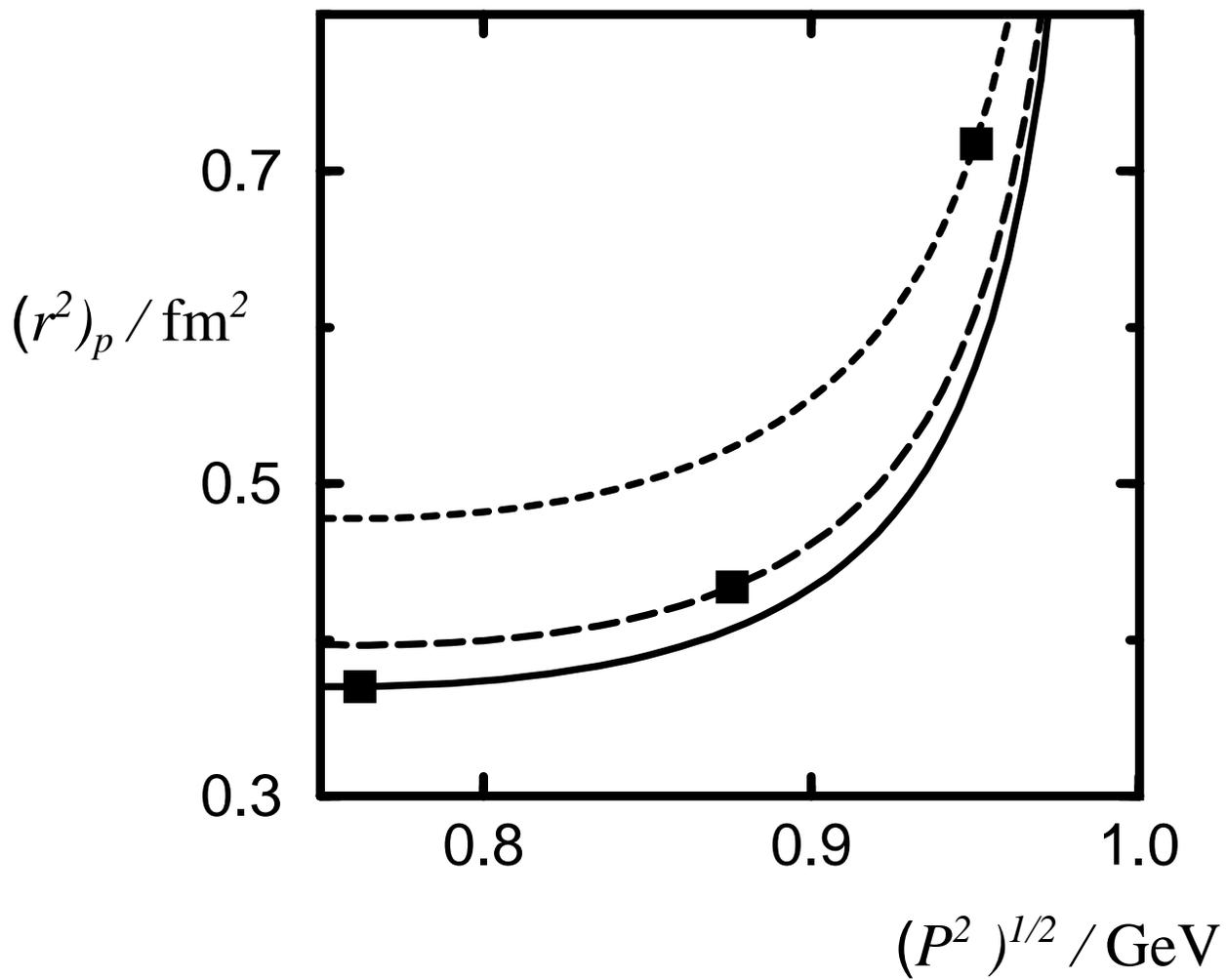
{\bf Fig.\ 1}: {\it The dependence of the proton charge radius,
$\langle r^2 \rangle_p$ on the proton c.o.m.\ momentum,
$P^2 \equiv \PPf^2 = \PPi^2 \neq m_B^2$. The charge radius is calculated
from {\rm eq.(\ref{Gamma_B_static})} for arbitrary $P^2$, with amplitudes
normalized according to {\rm eq.(\ref{normalization})}, but at
$P^2\neq m_B^2$. Shown are the radii obtained with
$\Lambda_B = 0.55\,{\rm GeV}$ (solid line), $0.45\,{\rm GeV}$ (dashed line)
and $0.35\,{\rm GeV}$ (dotted line).
The on--shell masses corresponding to the different values of $\Lambda_B$
are marked by squares on the respective curves. Here, for all $\Lambda_B$,
the constituent quark mass is $M = 0.4\,{\rm GeV}$, and for the diquark
mass we have chosen a fixed value of $m_d = 0.6\,{\rm GeV}$, so that the
diquark--quark continuum starts at $1.0\,{\rm GeV}$.}
\end{document}